\documentclass[aps,prl,reprint,showpacs,superscriptaddress,longbibliography]{revtex4-1}
\usepackage{mathtools,amssymb,graphicx,units}
\usepackage[usenames,dvipsnames]{color}
\usepackage[plainpages=false,pdfpagelabels,colorlinks=true,linkcolor=red,urlcolor=blue,citecolor=blue,pdftitle={Title},pdfauthor={},pdfdisplaydoctitle=true,pdfduplex=DuplexFlipLongEdge]{hyperref}

\usepackage{subfigure}
\usepackage{grffile}
\usepackage{bm}
\usepackage{mhchem}
\usepackage{color}
\usepackage{hyperref}
\usepackage{bibentry}

\newcommand{\eval}[3]{\langle#1\vert#2\vert#3\rangle}

\newcommand{\ud}{\mathrm{d}}

\newcommand{\ignore}[1]{}
\newcommand{\nobibentry}[1]{{\let\nocite\ignore\bibentry{#1}}}
\newcommand{\bibfnamefont}[1]{#1}
\newcommand{\bibnamefont}[1]{#1}

\begin{document}

\title{Photoinduced enhancement of bond order in the one-dimensional extended Hubbard model}

\author{Can Shao}
\email{shaocan2018@csrc.ac.cn}
\affiliation{Beijing Computational Science Research Center, Beijing 100084, China}

\author{Hantao Lu}
\email{luht@lzu.edu.cn}
\affiliation{Center for Interdisciplinary Studies $\&$ Key Laboratory for Magnetism and Magnetic Materials of the MoE, Lanzhou University, Lanzhou 730000, China}

\author{Hong-Gang Luo}
\affiliation{Center for Interdisciplinary Studies $\&$ Key Laboratory for Magnetism and Magnetic Materials of the MoE, Lanzhou University, Lanzhou 730000, China}
\affiliation{Beijing Computational Science Research Center, Beijing 100084, China}

\author{Rubem Mondaini}
\email{rmondaini@csrc.ac.cn}
\affiliation{Beijing Computational Science Research Center, Beijing 100084, China}


\begin{abstract}
We investigate the real-time dynamics of the half-filled one-dimensional extended Hubbard model in the strong-coupling regime, when driven by a transient laser pulse. Starting from a wide regime displaying a charge-density wave in equilibrium, a robust photoinduced in-gap state appears in the optical conductivity, depending on the parameters of the pulse. Here, by tuning its conditions, we maximize the overlap of the time-evolving wavefunction with excited states displaying the elusive bond-ordered wave of this model. Finally, we make a clear connection between the emergence of this order and the formation of the aforementioned in-gap state, suggesting the potential observation of purely electronic (i.e., not associated with a Peierls instability) bond-ordered waves in experiments involving molecular crystals.
\end{abstract}


\maketitle

\paragraph{Introduction.} \label{sec1}
Driving nonequilibrium behavior in strongly interacting systems has been used as a way to unveil singular information about the different degrees of freedom that give rise to their ordered phases. A clear paradigm of this scenario is given in the context of optical excitations in pump-probe experiments, where one is able to transiently induce ultrafast transitions between different electronic phases, as a result of tuning either their structural, magnetic, or electronic properties~\cite{Cavalleri17_review, Cavalleri18_review, Wang2018}. 
Their specific nature depends on the characteristics of the pump pulse and on the material under investigation. For example, it is possible to induce or enhance superconductivity at short time scales if melting some of their competing orders, as the static charge stripes that appear at optimally underdoped cuprates~\cite{Fausti11, Nicoletti14, Hu14, Forst14}.

In other situations, magnetic~\cite{Forst11, Graves13} or insulator-to-metal~\cite{Iwai03, Oka03, Oka05, Okamoto07, Takahashi08, Eckstein10} transitions are accomplished either by driving with a strong electric field or with a transient laser pulse. All these achievements largely rely on the development in the past few decades of ultrafast techniques, such as transient transmissivity (reflectivity) spectroscopy measurements, from which time-resolved optical conductivity can be extracted via Kramers-Kronig transformations~\cite{Roessler65}. A sub set of these studies concerns materials where, due to their peculiar crystal structure, one-dimensional (1D) models are believed to capture the nature of their electronic phases. In particular, molecular solids, as the bis(ethylendithyo)-tetrathiafulvalene-difluoro-tetracyanoquinodimethane (ET-$\text{F}_2$TCNQ), are viewed as good examples of 1D Mott insulators whose chains formed by ET molecules possess large on-site and nearest-neighbor (NN) Coulomb repulsions, resulting in electronic immobility~\cite{Hasegawa00}. Others, as some halogen-bridged compounds, are illustrative of charge-density-wave (CDW) insulators~\cite{Matsuzaki06}.

In both cases, the simplest model potentially describing their equilibrium properties is the extended Hubbard model (EHM), written as
\begin{eqnarray}
\hat H=-&t_h&\sum_{i,\sigma}\left(\hat c^{\dagger}_{i,\sigma} \hat c^{\phantom{}}_{i+1,\sigma}+\text{H.c.}\right)+U\sum_{i}\hat n_{i,\uparrow}\hat n_{i,\downarrow} \nonumber \\
+&V&\sum_{i}\hat n_{i}\hat n_{i+1},
\label{eq:H}
\end{eqnarray}
where $\hat c^{\dagger}_{i,\sigma}$ ($\hat c^{\phantom{}}_{i,\sigma}$) is the creation (annihilation) operator of an electron with spin $\sigma$ at site $i$, and the number operator is $\hat n_i= \hat n_{i,\uparrow}+ \hat n_{i,\downarrow}$; $t_h$ denotes the hopping amplitude, while $U$ and $V$ the on-site and NN Coulomb repulsions, respectively.

Ultrafast photoirradiation of these materials has revealed unique out-of-equilibrium responses, as the induction of transient metallic behavior~\cite{Iwai03, Okamoto07, Uemura08}, generation of insulating behavior with different characteristics, as from Mott-to-CDW insulators~\cite{Matsuzaki14}, and the change in the nature of charge orders~\cite{Matsuzaki03}. In other recent pump-probe measurements of the organic Mott insulator ET-$\text{F}_2$TCNQ, a new resonance appears after photoexcitation and implies the manifestation of an in-gap state~\cite{Wall2010}, also observed in theoretical analyses~\cite{Lu15}.
This state is attributed to the electronic delocalization through quantum interference between bound and ionized holon-doublon pairs, transiently induced by the pulse.

The ground-state (GS) phase diagram of~\eqref{eq:H} displays phases where the on-site and NN Coulomb interactions compete so as to induce insulating behavior with either spin-density-wave (SDW) or charge-density-wave (CDW) orders at large $U$ and $V$, respectively, connected via a first-order phase transition at $U=2V$~\cite{Dongen94}. At smaller values of the interactions, however, an elusive intermediate bond-ordered-wave (BOW) phase has been demonstrated~\cite{Nakamura99, Nakamura00, Jeckelmann02, Sengupta02, Tsuchiizu02, Sandvik04, Zhang04, Aichhorn04, Ejima07}. Our main result in this Rapid Communication is to argue on the possible observation of in-gap states at the out-of-equilibrium optical conductivity precisely associated with the induction of a BOW phase in a parent equilibrium regime displaying CDW order.

\paragraph{Methods and observables.}
We focus on the zero-temperature strong-coupling regime, with $U=10$ --- hereafter, we set the energy scale $t_h=1$ --- which is consistent with the estimated on-site interaction of ET-$\text{F}_2$TCNQ materials~\cite{Hasegawa00}. In theory, a first-order phase transition between the SDW and CDW GS's occurs at $V \sim 5$, sufficiently far from possible influences of the dimerized BOW phase, which is believed to exist up until a critical point at smaller interaction strengths, $(U_c, V_c) = (9.25, 4.76)$~\cite{Sandvik04, Ejima07}.

The system, when driven out of equilibrium by a transient pumping pulse, is affected by a time-dependent electric field (vector potential), whose introduction is done via the Peierls' substitution,
\begin{equation}
\hat c^{\dagger}_{i,\sigma}\hat c_{i+1,\sigma}+\text{H.c.}\rightarrow
e^{\mathrm{i}A(t)}\hat c^{\dagger}_{i,\sigma}\hat c_{i+1,\sigma}+\text{H.c.}.
\label{eq:Peierls}
\end{equation}
In terms of the temporal gauge, the vector potential in \eqref{eq:Peierls} is written as
$A(t)=A_0e^{-\left(t-t_0\right)^2/2t_d^2}\cos\left[\omega_0\left(t-t_0\right)\right]$, i.e., its temporal distribution is Gaussian centered around $t_0$, with $t_d$ controlling its width, and $\omega_0$ the frequency~\cite{Hashimoto14, Hashimoto15, Hashimoto17, Wang16, Wang17, Wang18, Shinjo19}. We use short-lived pulses by selecting $t_d=0.5$ (in terms of the time unit, $t_h^{-1}$) so as to describe the dynamics of ultrafast irradiations.

By employing the time-dependent Lanczos method, the evolved wave function $|\psi(t)\rangle$ can be computed starting from the initial GS $|\Psi_0\rangle$~\cite{Manmana2005,Prelovsek,SM}. To mitigate the influence of finite-size effects in our lattices of length $L$, we further contrast our results with the application of a twisted boundary condition (TBC) averaging~\cite{Poilblanc91, Tohyama04}, where the Peierls substitution~\eqref{eq:Peierls} acquires an extra phase $e^{\mathrm{i}A(t)}\hat c^{\dagger}_{i,\sigma}\hat c_{i+1,\sigma}+\text{H.c.}\to e^{\mathrm{i}A(t)}e^{\mathrm{i} \kappa}\hat c^{\dagger}_{i,\sigma}\hat c_{i+1,\sigma}+\text{H.c.}$, with $\kappa=\phi/L$ and $\phi\in[0,2\pi)$, enabling the evolution from the $\kappa$-dependent initial state $|\Psi_0^{\kappa}\rangle$ to $|\psi^{\kappa}(t)\rangle$.

The transport in this strongly interacting system can be quantified by the optical conductivity $\sigma(\omega)$, which in equilibrium is given in terms of the Kubo formula~\cite{Mahan}. While there is no well-defined out-of-equilibrium optical conductivity, because of the absence of time translation invariance, various methods to calculate $\sigma(\omega)$ in and out of equilibrium, as well as their validity in different limits, have been demonstrated in Ref.~\onlinecite{Shao16}. Here, we adopt the method derived rigorously from linear-response theory~\cite{Zala14},
\begin{equation}
\sigma(\omega,t)=\int_0^{t_m}\sigma(t+s,t)e^{{\mathrm i}(\omega+i\eta)s}\,\ud s,
\label{eq:sigmaomega}
\end{equation}
\begin{equation}
\sigma(t',t)=\frac{1}{L}\left[\eval{\psi(t')}{\hat{\tau}}{\psi(t')}+\int_t^{t'}\chi(t',t'')\,\ud t''\right],
\label{eq:sigmatt}
\end{equation}
where the two-time susceptibility is
\begin{equation}
	\chi(t',t'')=-\mathrm{i}\theta(t'-t'')\eval{\psi(t)}{[\hat \jmath^{I}(t'),\hat \jmath^{I}(t'')]}{\psi(t)},
\label{eq:chi}
\end{equation}
and in the diamagnetic term, the stress tensor operator reads $\hat \tau=t_h\sum_{i,\sigma}(\hat c_{i+1,\sigma}^{\dagger}\hat c_{i,\sigma}+\text{H.c.})$. The maximum time $t_m$ for the Fourier transformation [Eq.~\eqref{eq:sigmaomega}] in our numerical simulation is $\sim 100\,t_h^{-1}$. The interaction representation of the current operator $\hat \jmath ^{I}(t^\prime)$ is defined as $U^\dagger(t^\prime,t)\ \hat\jmath \ U(t^\prime,t)$, where $U(t^\prime,t)$ is the time-evolution operator {\em in the absence of probing perturbations}~\cite{Zala14}. Lastly, the current operator reads $\hat \jmath=-\mathrm{i} t_h\sum_{i,\sigma}[\hat c_{i,\sigma}^{\dagger}\hat c_{i+1,\sigma}-\text{H.c.}]$.

In what follows, we define the time difference between the pump's central time and the probe time as $\Delta t$, finally tracking $\sigma(\omega, \Delta t)$, intimately connected to the time-dependent reflectivity in experiments.

\paragraph{Results.}
\begin{figure}[ht]
\centering
\includegraphics[width=0.5\textwidth,height=0.3\textheight]{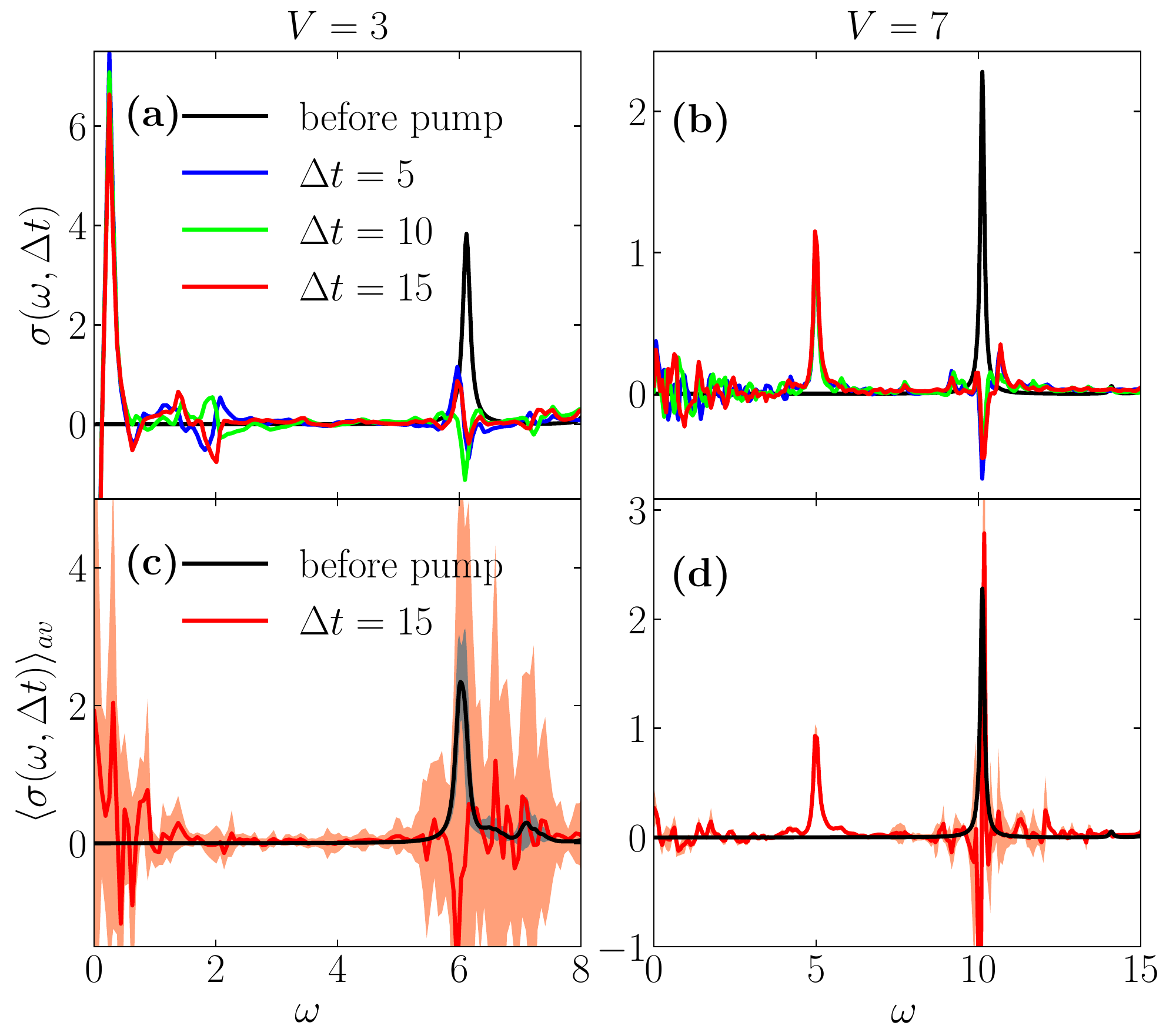}
\caption{
Re$\sigma(\omega, \Delta t)$ for a lattice with $L=14$ and $U=10$.
The standard periodic BC (TBC averaging) is used in (a) [(c)] with $V=3$ in SDW and (b)[(d)] with $V=7$ in CDW. The black (red) solid line in (c) and (d) is the averaged result before (after) the pump over ten equidifferent twisted phases $\phi\in[0,2\pi)$, with the shading marking the corresponding error bar. Parameters of the pump: $A_0=0.4$ and $t_d=0.5$, with $\omega_0$ matching the position of the main peak in equilibrium. The broadening factor $\eta$ is taken to be $1/L$.
}
\label{fig_OC}
\end{figure}

\begin{figure}[ht]
\centering
\includegraphics[width=0.5\textwidth,height=0.3\textheight]{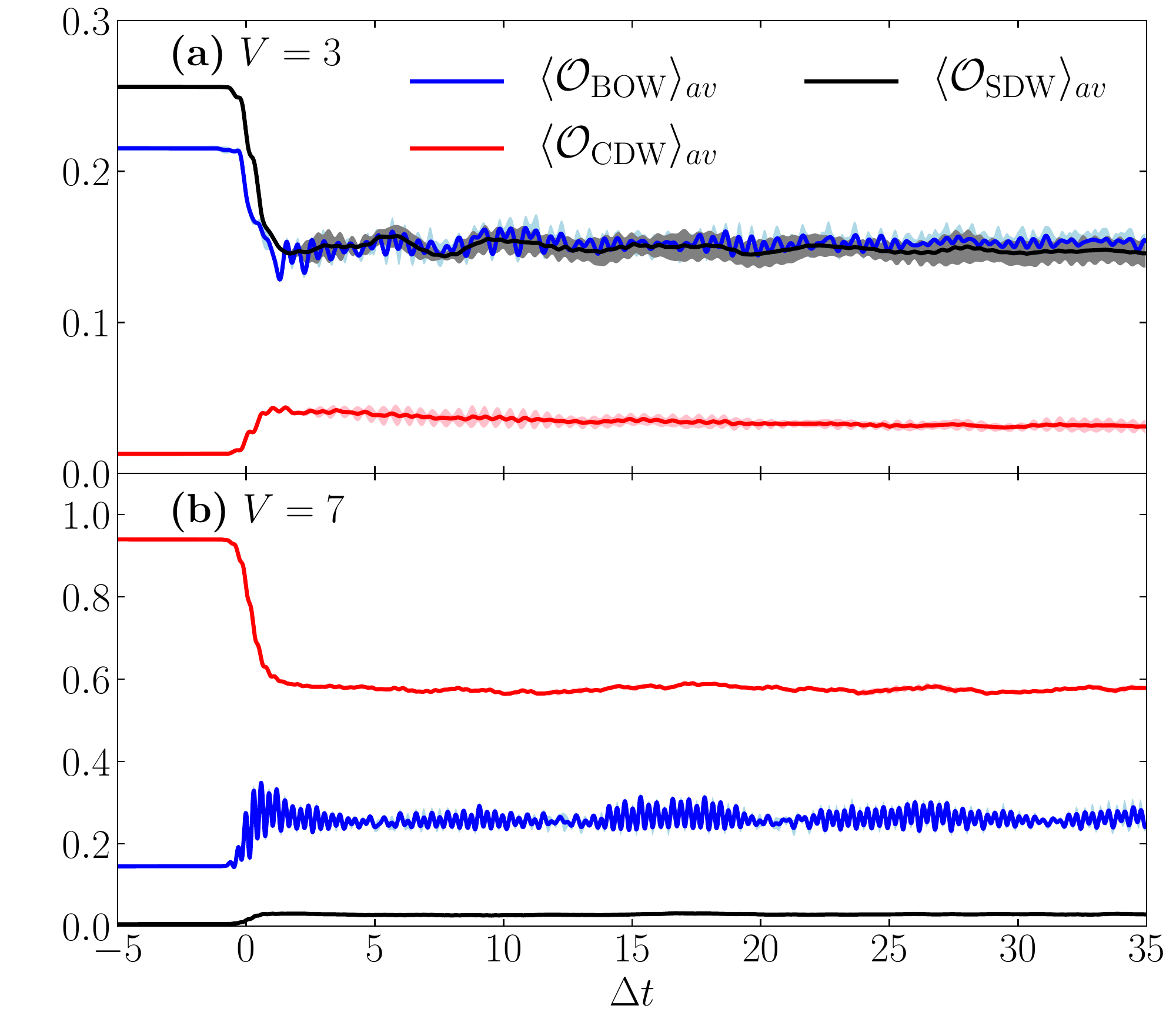}
\caption{
The time-dependent normalized structure factors of BOW (blue), CDW (red), and SDW (black) before and after the pump with $V = 3$ ($V = 7$) in (a) [(b)], for a lattice with $L=14$ and $U=10$. As before, the solid lines represent the averaged result over ten equidifferent twisted BCs with shading marking the error bar. Parameters of the pump: $A_0=0.4$, $t_d=0.5$, and $\omega_0$ matching the position of the main peak in the  equilibrium $\sigma(\omega)$.
}
\label{fig_order}
\end{figure}

We start by comparing the optical conductivity computed from GSs in each side of the transition, with $V = 3$ and $7$, symmetric with respect to the transition point $V\simeq5$ (for $U = 10$). We report its real part, Re $\sigma(\omega, \Delta t)$, in a lattice with $L=14$ and standard periodic boundary conditions (BCs), in Figs.~\ref{fig_OC}(a) and \ref{fig_OC}(b), respectively, both before (in equilibrium) and after the pump ($\Delta t=5$, $10$, $15$). The size of the optical gap, i.e., the position of the main peak in equilibrium, is $\omega_{\rm gap}\simeq6.12$ ($\simeq10.12$) for $V=3$ ($V=7$). To excite the system, we thus resonantly apply the pump, selecting $\omega_0 = \omega_{\rm gap}$, also setting $A_0 = 0.4$, so as to enhance the bond order as will later become clear. In both cases, the original peak at $\omega_{\rm gap}$ is suppressed after the pump, while another peak arises at smaller energies. We dub these photoinduced states below $\omega_{\rm gap}$ as the in-gap states, occurring at $\omega_{\text{in-gap}}$. For the situation initially displaying SDW order [Fig.~\ref{fig_OC}(a)], the in-gap peak is extremely close to $\omega=0$, suggesting it might be indeed zero when approaching $L\to\infty$. Figures~\ref{fig_OC}(c) and \ref{fig_OC}(d) display the same as in Figs. 1(a) and 1(b), but employing the TBC averaging with ten equidifferent twisted phases $\phi\in[0, 2\pi)$. Although still noisy for this system size, this induced peak at long times approaches $\omega=0$, indicating a metallic regime. In stark contrast, the in-gap state generated around $\omega\approx5$ for excitations from the CDW phase does not change regardless of time and TBCs [Fig.~\ref{fig_OC}(d)], which is indicative it may well exist in the thermodynamic limit.

\begin{figure}[ht]
\centering
\includegraphics[width=0.5\textwidth,height=0.3\textheight]{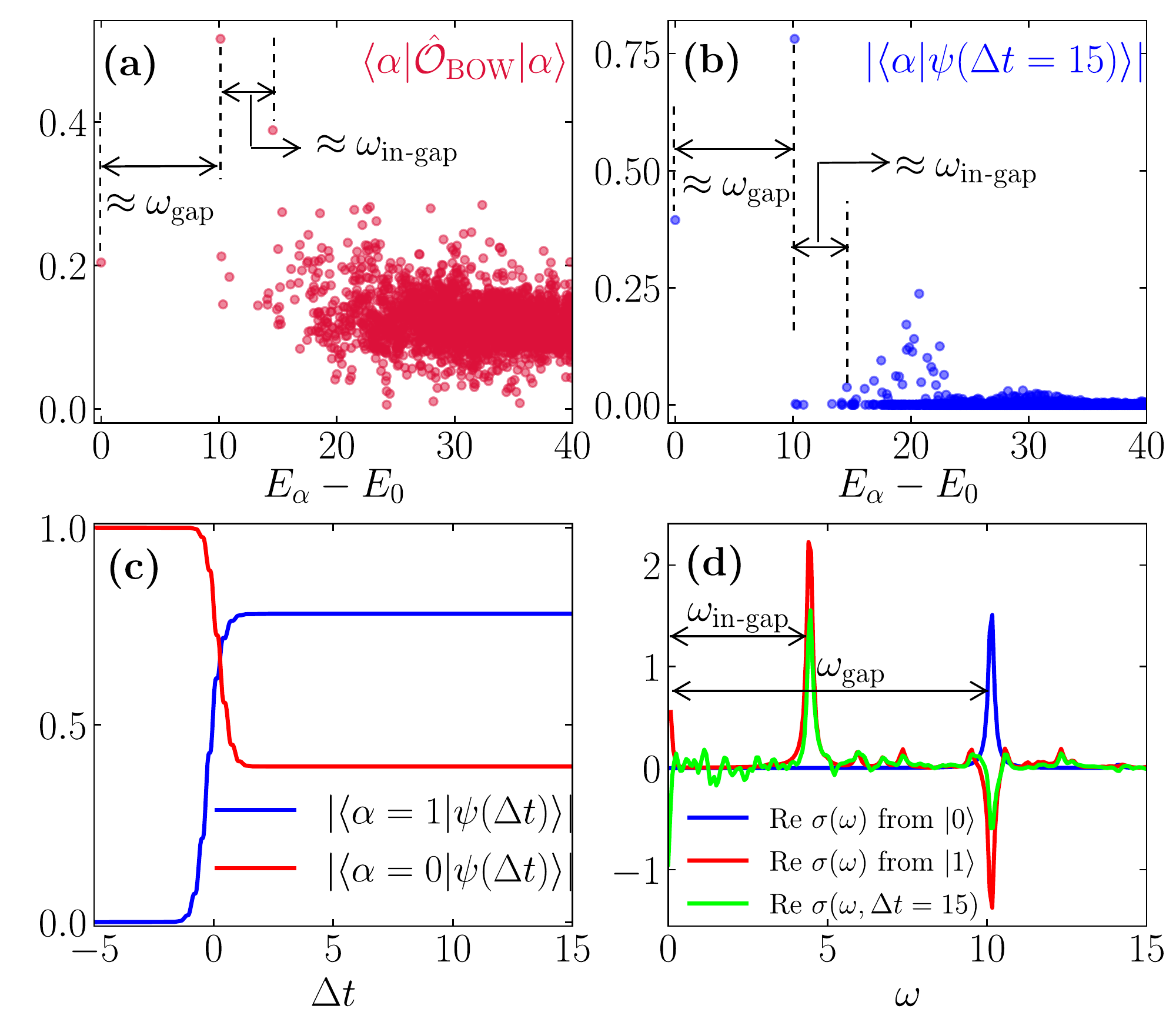}
\caption{Full-ED calculation, restricted to the $k=0$ subspace, of a ten-site lattice with parameters $U=10$ and $V=7$. $E_{\alpha}-E_0$ is the energy difference between the corresponding excited states $|\alpha\rangle$s and the GS. (a) $\langle\alpha|\hat{\mathcal{O}}_{\text{BOW}}|\alpha\rangle$, eigenstate expectation values of the normalized BOW structure factor vs $E_{\alpha}-E_0\in[0,40]$. (b) $|\langle\alpha|\psi(\Delta t=15)\rangle|$, overlap between $|\psi(\Delta t=15)\rangle$ and eigenstates $|\alpha\rangle$s vs $E_{\alpha}-E_0\in[0,40]$. (c) $|\langle\alpha|\psi(\Delta t)\rangle|$, overlap between $|\psi(\Delta t)\rangle$ and the eigenstates $|\alpha=0,1\rangle$ of the equilibrium Hamiltonian vs $\Delta t$. (d) $\sigma(\omega)$ calculated from the GS (blue) and first excited state (red), as well as $\sigma(\omega, \Delta t=15)$ (green) lines. Parameters of the pump: $A_0=0.5$, $t_d=0.5$, and $\omega_0=10.12$ matches the optical gap.
}
\label{fig_pump}
\end{figure}
The question now boils down to understanding the physical nature of the photoinduced in-gap state generated in the CDW regime. For that purpose, we recall the different structure factors associated with the three different insulating phases observed in equilibrium in the case of repulsive interactions: SDW, CDW, and BOW. We generically define those in a translationally invariant and staggered fashion as
\begin{equation}
\mathcal{\hat O}_{x} = \frac{1}{L^2}\sum\limits_{i=0}^{L-1}\sum\limits_{dx=0}^{L-1}{(-1)}^{dx} \hat O_i\hat O_{i+dx},
\end{equation}
with $\hat O_i = \hat n_{i,\uparrow}-\hat n_{i,\downarrow}$, for $x= \text{SDW}$; $\hat O_i = \hat n_{i}$, for $x = \text{CDW}$ and $\hat O_i =  \sum_{\sigma}(\hat c^{\dag}_{i}\hat c_{i+1}+ \text{H.c.})$, for $x = \text{BOW}$; $dx$ represents the distance from site $i$~\footnote{It is worth stressing that the definition of the order parameters in Refs. [\onlinecite{Tsuchiizu02}] and [\onlinecite{Ejima07}] is only suitable for open BCs, while in the case of periodic BCs, with translation invariance, the only possible way to define them is via the summation over correlation functions, as shown in the main text.}. For the last two phases, the $L\to\infty$ extrapolation of this quantity corresponds to the square of the corresponding order parameters.

In Fig.~\ref{fig_order}, we show the BC averaged time evolution of these three normalized structure factors, $\langle {\cal O}_{x}\rangle_{\rm av} \equiv (1/N_\kappa)\sum_\kappa\langle \psi^\kappa(t)| {\cal \hat O}_{x}|\psi^\kappa(t)\rangle$, using $N_\kappa$ equidifferent TBCs. For $V=3$ [Fig.~\ref{fig_order}(a)], the pump is responsible for inducing a metallic behavior as indicated by the optical conductivity peaks. This happens at the expense of substantially reducing the SDW correlations. Conversely, the CDW and BOW structure factors are slightly changed with extremely long saturation times. A proper finite-size scaling would rule out the manifestation of any order in the thermodynamic limit, but given the metallic behavior suggested by $\sigma(\omega,\Delta t=15)$, one would not expect their concomitant appearance. On the other hand, when $V=7$ [Fig.~\ref{fig_order}(b)], there is a considerable increment of the BOW order with little influence of the  different TBCs (the shadings are barely visible), at the cost of a dramatic reduction of the ruling order parameter in equilibrium, proportional to $\cal O_{\rm CDW}$.

Given this enhancement of the BOW structure factors, we are now in a position to correlate the appearance of the in-gap state with a photoinduced bond order. To verify this point, we perform a full exact diagonalization (ED) calculation in a ten-site lattice, restricted to the $k=0$ momentum subspace. This is the sector where the equilibrium ground state resides and where the time-evolved wave function explores, since the pump does not break translation invariance. Figure~\ref{fig_pump}(a) displays the eigenstate expectation values of the BOW structure factors, for eigenstates $|\alpha\rangle$'s of the equilibrium Hamiltonian, as a function of the energy difference $E_{\alpha}-E_0$, where $E_0$ is the GS energy. One finds that the first excited state ($E_{1}-E_0\simeq10.13$) displays the largest bond order ($\langle 1|\mathcal{O}_{\text{BOW}}|1\rangle \simeq 0.52$) among all $|\alpha\rangle$'s. Besides, Fig.~\ref{fig_pump}~(b) shows the overlap between the evolved wave function at long times after the pump, and all eigenstates, i.e., $|\langle\alpha|\psi(\Delta t=15)\rangle|$. The overlap with $\alpha=1$ reaches values up to $0.783$ with the optimal pump parameters: $A_0=0.5$ and $\omega_0=10.12$. The detailed time evolution of the overlap between $|\psi(\Delta t)\rangle$ with both the GS and the first excited state is shown in Fig.~\ref{fig_pump} (c). Their weight switch roles, as the pump reaches its maximum intensity at $\Delta t=0$. Notice as well that $E_1-E_0\simeq10.13$ is consistent with the optical gap $\omega_{\rm gap}\simeq10.12$, indicative that the system displays a large resonance so as to absorb energy sufficient to excite this $|\alpha=1\rangle$ state~\footnote{One can easily understand this argument directly from the definition of the Kubo formula, provided the optical conductivity is composed of one large single-peak.}.

\begin{figure}[ht]
\centering
\includegraphics[width=0.5\textwidth,height=0.30\textheight]{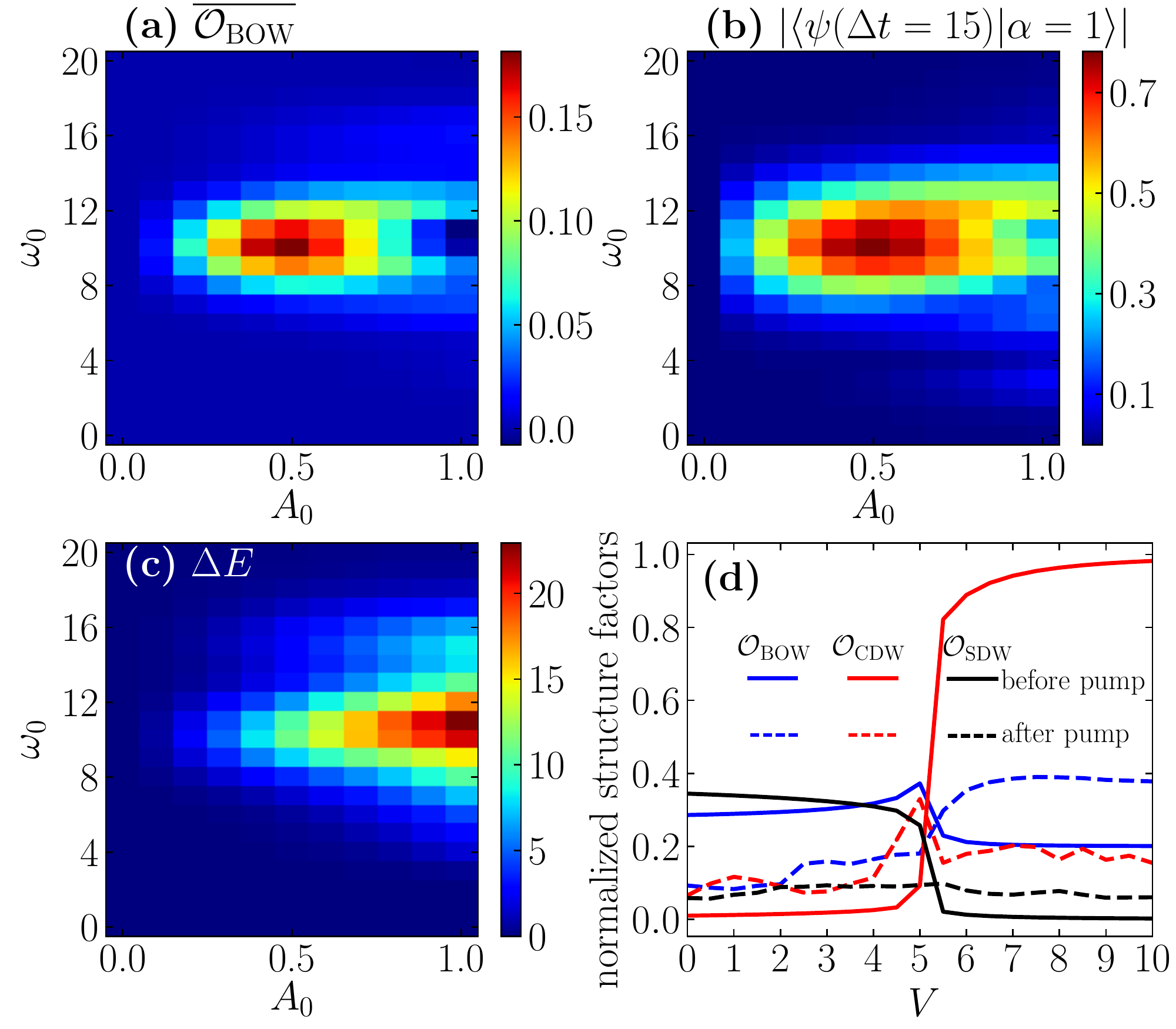}
\caption{
Contour plots of a long time-evolution increase of the BOW order $\overline{{\cal O}_{\rm BOW}}$ obtained by averaging on interval $\Delta t\in[5,105]$ in (a); overlap between wave function $|\psi(\Delta t=15)\rangle$ and the first excited state $|1\rangle$ in (b); and (c) the injected energy $\Delta E$, all as a function of $A_0$ and $\omega_0$ with $V=7$.
(d) The normalized structure factors of BOW, CDW, and SDW before (solid lines) and after the pump (dashed lines) at very long times, with $A_0\in[0,1]$ and $\omega_0\in[0,20]$, as a function of $V$. The lattice size is $L=10$ and the pump time width is $t_d=0.5$, in a mesh of $A_0$ and $\omega_0$ values with discretization 0.1 and 1.0, respectively.
}
\label{fig_ingap}
\end{figure}

To finally confirm the relation between the pump-enhanced bond order and the in-gap state observed in the optical conductivity displayed in Fig.~\ref{fig_OC}(b), we show in Fig.~\ref{fig_pump}(d) the time-evolved $\sigma(\omega, \Delta t=15)$ and the equilibrium $\sigma(\omega)$ computed from the GS [as in Fig.~\ref{fig_OC} (b)], accompanied by the equilibrium optical conductivity computed from the first excited state. The similarity between $\sigma(\omega, \Delta t=15)$ and $\sigma(\omega)$ from $|1\rangle$ makes clear the nature of the in-gap state: It is related to a photoinduced bond order. What is more interesting is that the energy associated with the in-gap state is precisely the energy difference between the first excited state and the state with the second largest $\langle \hat {\cal O}_{\rm BOW}\rangle$ in the eigenspectrum [see the annotation in Fig.~\ref{fig_pump}(a)]. This indicates that not only the pump results in a state displaying BOW order, but also connects its first excitation to states displaying the same symmetry.

The final point we address is in systematically finding the optimal parameters of the pump that leads to a BOW enhancement. In Figs.~\ref{fig_ingap}(a)-\ref{fig_ingap}(c), we give the contour plots of the normalized BOW structure factor at long times, $\overline{\mathcal{O}_{\rm BOW}}$, the overlap of $|\langle\alpha=1|\psi(\Delta t=15)\rangle|$, and the injected energy $\Delta E\equiv \langle\psi(t)|\hat H|\psi(t)\rangle - \langle\Psi_0|\hat H|\Psi_0\rangle$, as a function of pump parameters $A_0$ and $\omega_0$,  with $V=7$, respectively. Here, we do not use the twisted BCs because their influence in these quantities  is small [see Fig.~\ref{fig_order}(b)]. The optimal $\omega_0$ precisely coincides with $\omega_{\rm gap}$ and as Fig.~ \ref{fig_ingap}(c) shows, the system absorbs more energy if $\omega_0$ is closer to $\omega_{\rm gap}$, as one varies $A_0$. Lastly, the overlap of the wave function at long times and the first excited state in Fig.~\ref{fig_ingap}(b) displays a remarkable similarity with Fig.~\ref{fig_ingap}(a), confirming the connection between the enhanced BOW order and the overlap increase between the time-evolved wave function and the first excited state. A detailed analysis on the dependence with the pump parameters $(A_0, t_d)$ is presented in the Supplemental Material~\cite{SM}.

As a final remark on the generality of our results, Fig.~\ref{fig_ingap}(d) contrasts the equilibrium (before pump) structure factors of the three phases we investigate (solid lines) and $\overline{\mathcal{O}_{x}}$, i.e., the long-time average (obtained within $\Delta t \in [5, 105]$) for each of the structure factors (dashed lines), always optimizing the pump variables $A_0$ and $\omega_0$ (with $A_0\in[0,1]$ and $\omega_0\in[0,20]$) such as to enhance the corresponding order, as a function of $V$.
The small enhancement of CDW order in the immediacy of the first-order phase transition in the SDW side (at $V\simeq4.5$) has been discussed in Ref.~\onlinecite{Lu12}. Remarkably, the enhancement of the BOW order within the equilibrium CDW phase is robust for a wide range of interactions $V$. Besides, we have further checked that the first excited state in this parameter space displays long-ranged BOW order~\cite{SM}.

\paragraph{Summary and discussion.} By utilizing the time-dependent Lanczos technique, we calculate the non-equilibrium optical conductivity and order parameters for different phases of the 1D EHM. We find that an enhancement of a BOW state can be readily reached from the GS of the equilibrium CDW phase of the model, when tuning the parameters of the pump so as to (i) be resonant with the main peak of the optical conductivity and (ii) with enough energy to induce a large overlap of the time-evolved wave function with the first excited state. We argue that in the background of alternating doublons and holons, the bond (dimerization) of electrons among the double occupied sites and their nearest empty site is one of the lowest-order excitations, which, under appropriate photoexcitation, can be dynamically accessed. This provides an unique framework for the observation of the elusive BOW order in experiments involving molecular crystals under ultrafast photoirradiation. Fundamentally, our emergent dimerization is intrinsic and not associated with electron-lattice couplings as observed in alkali-TCNQ compounds~\cite{McQueen_09, Uemura_12}.

\begin{acknowledgments}
C.S. and R.M. acknowledge support from NSAF-U1530401. R.M. also acknowledges support from the National Natural Science Foundation of China (NSFC) Grants No. 11674021  and No. 11851110757. H.T.L. and H.G.L. acknowledge support from NSFC Grants No. 11474136, No. 11674139, and No. 11834005, and the Fundamental Research Funds for the Central Universities. R.M. acknowledges discussions with C. Cheng; C.S. and H.L. acknowledge T. Tohyama for interactions in related works. The computations were performed in the Tianhe-2JK at the Beijing Computational Science Research Center (CSRC).\\
\end{acknowledgments}

\bibliography{lt}

\appendix
\section{Supplemental Material}
In this Supplemental Material, we highlight side aspects that complement the main message of photoinduced enhancement of bond-ordered wave (BOW), via ultrafast pumps, shown in the main text. We describe details of the numerical methods, the generality of the increase upon modifications of the pump characteristics, a discussion on non-coherent heating effects and, lastly, an analysis of the true long-range order in the first excited state within the charge-density wave equilibrium phase.

\paragraph{Time-dependent Lanczos method.} The time-dependent wave function is obtained under the unitary evolution promoted by $\hat H(t)$ [Eq.~(1) of the main paper] via a piecewise discretization of time, $|\psi(t+\delta{t})\rangle = e^{-{\rm i}\hat H(t)\delta t}|\psi(t)\rangle$, with time-stepping $\delta{t}$. For that, we employ a time-dependent Lanczos method, starting from the initial state given by the ground state of the corresponding equilibrium Hamiltonian in the absence of pump. In that approach, the evolution is given by [\href{https://aip.scitation.org/doi/abs/10.1063/1.2080353}{41}, \href{https://link.springer.com/chapter/10.1007/978-3-642-35106-8_1}{42}],
\begin{equation}
|\psi(t+\delta{t})\rangle\simeq\sum_{l=1}^{M}{e^{-{\mathrm i}\epsilon_l\delta{t}}}|\varphi_l\rangle\langle\varphi_l|\psi(t)\rangle,
\label{eq:lanczos}
\end{equation}
where $\epsilon_l$ and $|\varphi_l\rangle$ are eigenvalues and eigenvectors of $\hat{H}(t)$, respectively, in the corresponding Krylov subspace generated in the Lanczos iteration at each instant of time; $M$ is the dimension of the Lanczos basis. For the results presented, we selected $M=30$ and $\delta{t}=0.02t_h^{-1}$, where we have checked that within $t\leq200\,t_h^{-1}$, increasing the number of states $M$ does not produce substantial quantitative changes in our results for this $\delta t$.

\paragraph{BOW enhancement under different pump parameters.}
\begin{figure}[t]
\centering
\includegraphics[width=0.5\textwidth]{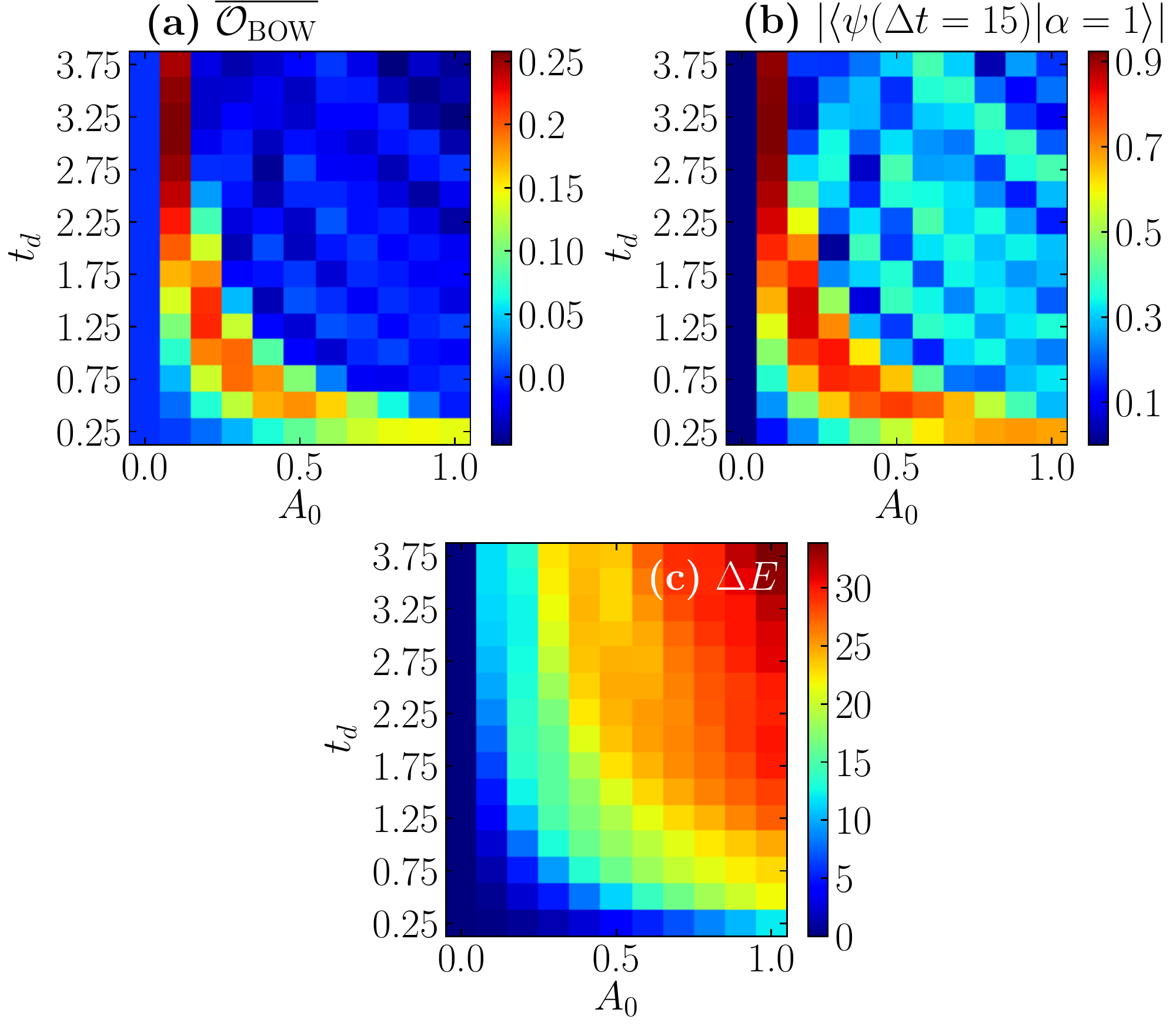}
\caption{
Contour plots of long time-evolution increase of the BOW order $\overline{{O}_{\rm BOW}}$ obtained by averaging on interval $\Delta t\in[5,105]$ in (a); overlap between wave function $|\psi(\Delta t=15)$ and the first excited state $|1\rangle$ in (b); and (c) the injected energy $\Delta E$, all as a function of $A_0$ and $t_d$ with $V=7$, and $U=10$.
}
\label{fig:A0td}
\end{figure}

An important aspect of our main result is its robustness under variations of the specific parameters of the pulse, namely its amplitude $A_0$ and duration $t_d$. In Fig.~4 of the main text, we show the influence of  $A_0$ and the pulse frequency $\omega_0$ on the enhancement of BOW order, intimately connected to the increase of the overlap of the time-evolved wave function with the first excited state. Now, by fixing the pump frequency $\omega_0 = 10.12$, resonant with the optical gap, we investigate the dependence of the pulse length in these results. Figure~\ref{fig:A0td} presents such analysis, for a lattice with $L=10$ and parameters $(U,V) = (10, 7)$. To start, in Fig.~\ref{fig:A0td}(a), the contour plot of the long time-evolution for the normalized BOW structure factor shows that $A_0$ and $t_d$ are intertwined: To obtain an enhancement for longer pulses one has to systematically reduce its amplitudes. Nonetheless, the regimes where the enhancement is achieved correspond to a wide ranges of  pulse parameters. As in the main text, this increase, also observed for different $t_d$'s, is directly connected to a large overlap of the time evolving wave function (in this case, obtained at $\Delta t =15$) with the first excited state $|\alpha=1\rangle$ of the equilibrium Hamiltonian, as shown in Fig.~\ref{fig:A0td}(b); it reaches an overlap as large as 0.9 at long pulse durations. Lastly, we report the dependence of the injected energy $\Delta E\equiv \langle\psi(t)|\hat H|\psi(t)\rangle - \langle\Psi_0|\hat H|\Psi_0\rangle$ on the pulse parameters in Fig.~\ref{fig:A0td}(c). Comparing to the previous panels, it becomes clear that the BOW amplification is obtained at $\Delta E \approx 10$, closely matching the gap between the ground state and the first-excited state with the same momentum quantum number, for these values of $(U,V)$.

\begin{figure}[t]
\centering
\includegraphics[width=0.5\textwidth,height=0.15\textheight]{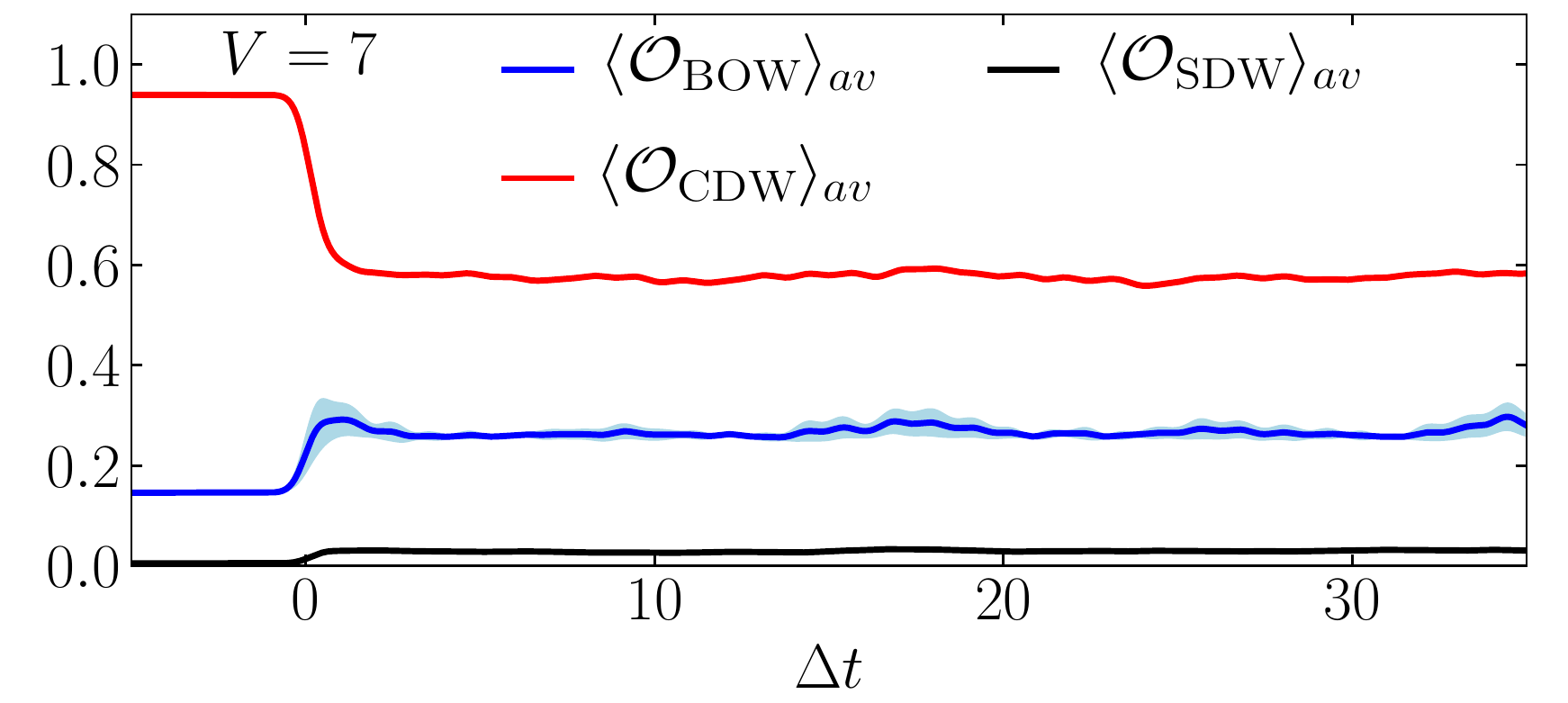}
\caption{
The time-dependent structure factors of BOW (blue), CDW (red) and SDW (black) before and after the pump with $L=14$, $U=10$ and  $V = 7$. The solid lines represent the averaged result over ten time phases $\varphi$ with shading marking the error bar. Parameters of the pump: $A_0=0.4$, $t_d=0.5$ and $\omega_0$ matching the position of the main peak of the optical conductivity in equilibrium $\sigma(\omega)$.
}
\label{fig:envelope}
\end{figure}

Concerning an experimental emulation closely related to our results, it is important to emphasize that due to finite time resolution, one cannot guarantee the maximum amplitude of the pump potential to precisely occur at $t_0$. As a consequence, it is important to analyze $A(t)$ with time  phases $\varphi$ as $A(t)=A_0e^{-\left(t-t_0\right)^2/2t_d^2}\cos\left[\omega_0\left(t-t_0\right)+\varphi\right]$ [\href{https://journals.aps.org/prb/abstract/10.1103/PhysRevB.96.235142}{38}]. Figure~\ref{fig:envelope} presents this analysis by checking the influence of the phases on the correspondent suppression or enhancement of the time-dependent structure factors, for the same conditions presented in Fig. 2(b) of the main text. By averaging them for ten different $\varphi$'s, we notice that only the coherent interaction-dependent oscillations in time are suppressed, but the overall quantitative aspects of their long-time averages are maintained.

\begin{figure}[t]
\centering
\includegraphics[width=0.49\textwidth]{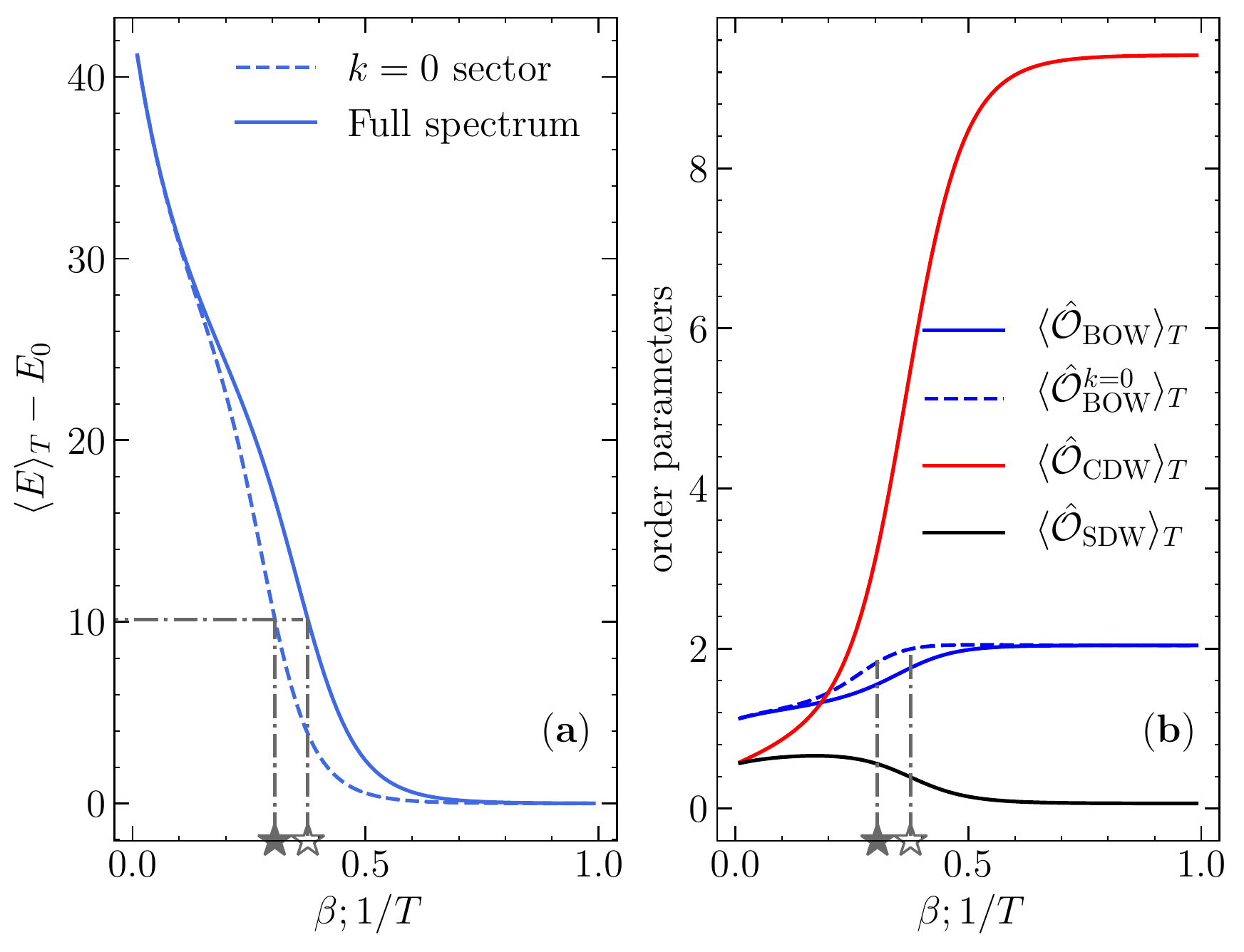}
\caption{
The temperature-dependent thermal energy (a) and thermal averaged structure factors (b) of BOW (blue), CDW (red) and SDW (black) with $L=10$, $U=10$ and $V = 7$. The solid and dashed line represent the results calculated using the full Hilbert space and $k=0$ subspace, respectively. (Empty) Full star markers signal the effective inverse temperature $\beta=1/T$ corresponding to the energy injected by the pump for the $k=0$ momentum sector (full Hilbert space), where $\beta=0.305$ (0.376).
}
\label{fig:thermal_effect}
\end{figure}

\paragraph{Finite temperature calculation.} Since the pump provides extra energy to the system, it is crucial to see whether our pulse-induced modification of the structure factors could be merely explained by the fact that its overall temperature has increased. This is called `heating effect' and would mask the coherent excitations the pulse may lead, which we claim to be fundamental to the BOW order enhancement. In our study, the system is isolated and its associated effective temperature $T$ can be inferred from the correspondent thermal mean energy (in units where $k_B=1$),
\begin{equation}
 \langle E\rangle_{T} = \frac{\sum_{\alpha}E_\alpha e^{-E_\alpha/T}}{\sum_\alpha e^{-E_\alpha/T}},
\end{equation}
provided one knows all the eigenenergies $E_\alpha$'s. For a system with 10 sites at half-filling, this is amenable and the only ambiguity comes from whether one selects either the momentum sector the wave-function explores or the full Hilbert space for the given total $S_z$ and number of electrons. In either case, the differences are shown to be small and decreasing with larger system sizes [\href{https://journals.aps.org/pre/abstract/10.1103/PhysRevE.81.036206}{54}]. With this, the effective temperature the system acquires after the pump is obtained via $\Delta E = \langle E\rangle_T - E_0$. For the pump parameters considered in the main text within the CDW phase $[(U,V) = (10,7)]$, $\Delta E \simeq 10.12$ and the corresponding (inverse) temperatures are signaled by the star markers in Fig.~\ref{fig:thermal_effect}(a). In a similar fashion, the thermal averages of the structure factors can also be obtained, provided one computes their eigenstate expectation values $\langle \alpha | \hat {\cal O}_{\rm x} |\alpha\rangle$. In that case, Fig.~\ref{fig:thermal_effect}(b) shows that for the effective temperature that corresponds to the amount of energy inserted by the pump, the thermal BOW
correlator decreases in comparison to the ground state. An effect direct at odds with the enhancement related to the coherent excitations promoted by $A(t)$. Therefore, our results under photoirradiation can not be explained as stemming from heating effects.

\begin{figure}[t]
\centering
\includegraphics[width=0.50\textwidth]{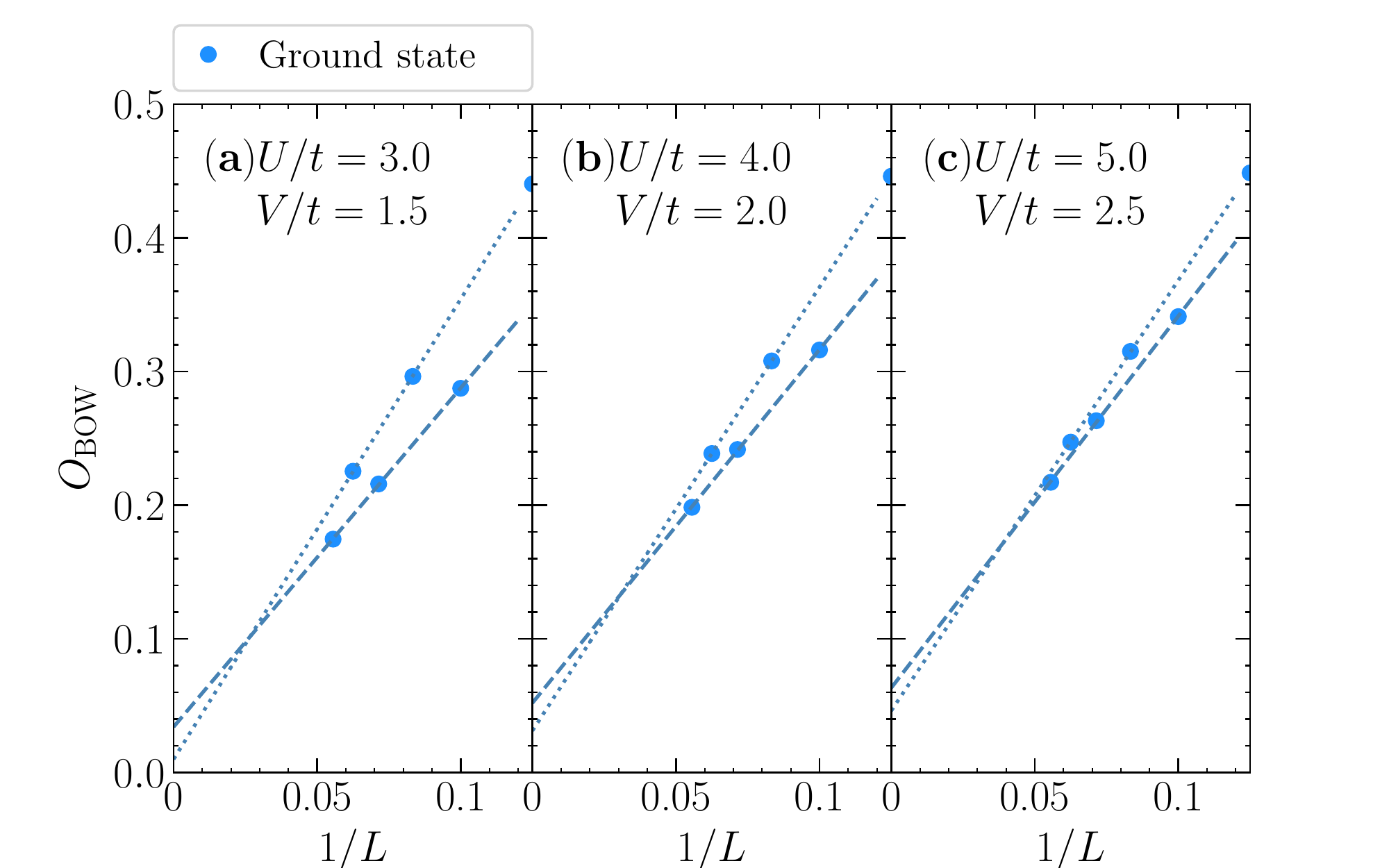}
\caption{
Finite size analysis of the BOW structure factor, defined as the normalized sum of all the staggered two-site correlation functions (see main text), for three points along the line $V=U/2$, known to display long range BOW order in the thermodynamic limit. Dashed (dotted) lines are linear fittings for the system sizes $L = 4n+2 \,\,(L = 4n)$, with integer $n$'s.
}
\label{fig:BOW_GS_scaling}
\end{figure}

\paragraph{Indications of long-range BOW order.} Both charge and spin gaps are finite in a BOW phase [\href{https://doi.org/10.1143/JPSJ.68.3123}{25}, \href{https://link.aps.org/doi/10.1103/PhysRevB.61.16377}{26}], which allows the ground state in this regime to display a full long-range order, even for the one-dimensional system considered [\href{https://link.aps.org/doi/10.1103/PhysRevB.65.155113}{28}, \href{https://link.aps.org/doi/10.1103/PhysRevLett.92.236401}{30}, \href{https://link.aps.org/doi/10.1103/PhysRevLett.99.216403}{33}]. Provided the accurate phase diagrams obtained in Refs. \href{https://link.aps.org/doi/10.1103/PhysRevLett.92.236401}{30} and \href{https://link.aps.org/doi/10.1103/PhysRevLett.99.216403}{33} for lattices up to $\sim1000$ sites, it is paramount to test whether the system sizes attainable by exact diagonalization methods are sufficient to observe a finite order parameter when approaching the thermodynamic limit. We show in Fig.~\ref{fig:BOW_GS_scaling} a finite size analysis along the line $V=U/2$, which resides within the BOW phase for values of $U \lesssim 9$, according to the most accurate phase diagram of this model obtained up to date [\href{https://link.aps.org/doi/10.1103/PhysRevLett.99.216403}{33}]. From our data, apart from the known ``even-odd'' effects for system sizes where $L/2$ is either odd or even in the presence of periodic boundary conditions, it is clear that using lattices as large as $L=18$ enables us to obtain a finite order parameter when approaching the thermodynamic limit.

\begin{figure}[t]

\centering
\includegraphics[width=0.47\textwidth]{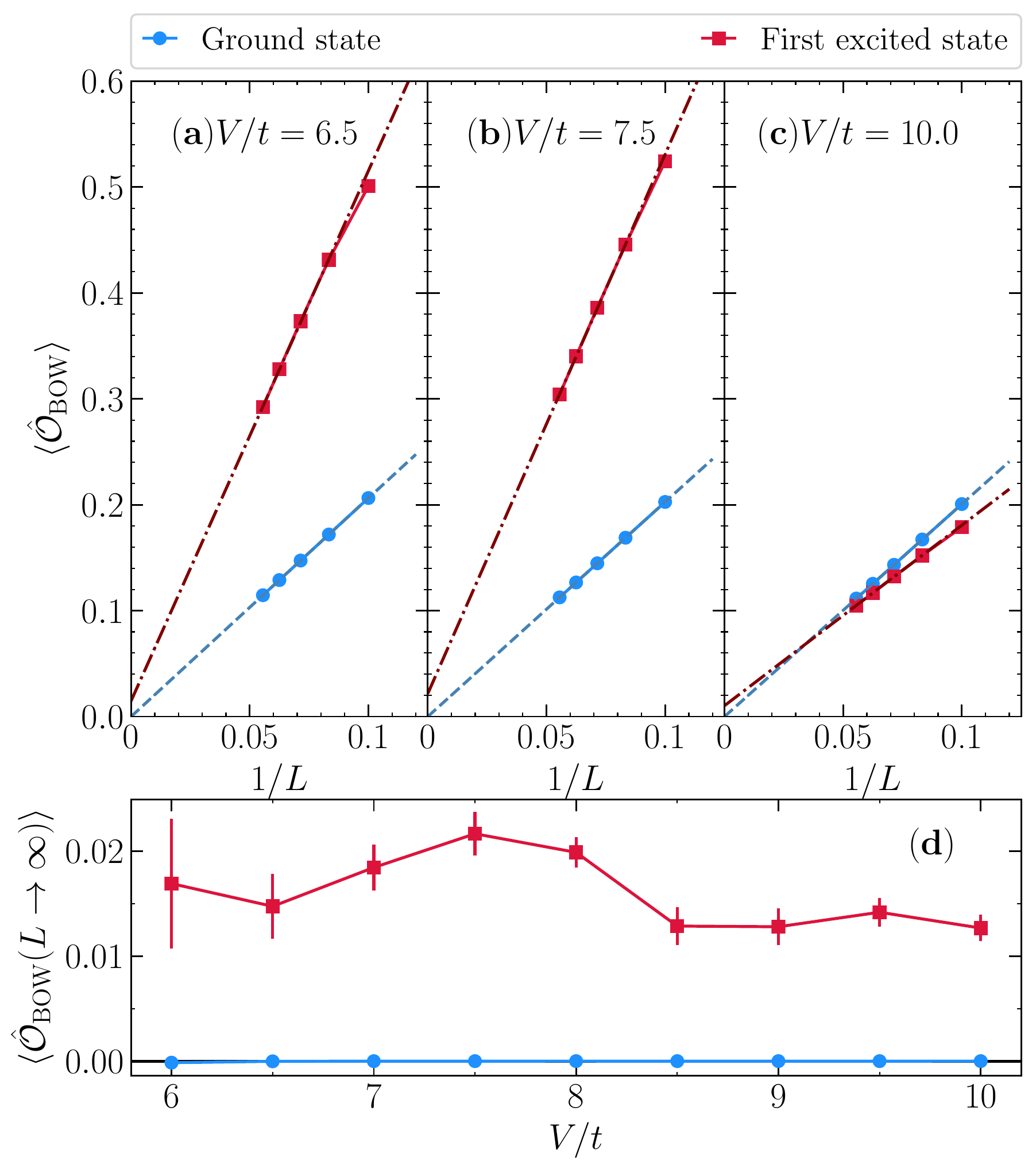}
\caption{
Finite size analysis of the BOW structure factor for fixed $U = 10$ and $V=6.5, 7.5$ and $10$ [(a), (b) and (c), respectively], performed for both the ground and the first excited states, with same momentum quantum numbers. The values when approaching the thermodynamic limit in the CDW phase are compiled in (d), and represent the corresponding square of the order parameters.
}
\label{fig:O_bow_scaling}
\end{figure}

Inspired by this result, in Fig.~\ref{fig:O_bow_scaling}, we apply the same analysis along the line $U = 10$ within the CDW phase, as primarily investigated in the main text. Due to the larger charge gaps in the strongly interacting regime, the convergence of the finite size results is also more accurate. First, we test whether the ground state displays a finite BOW order parameter when approaching the thermodynamic limit: the answer is negative, as one would expect, and the compilation of the $L\to\infty$ results are presented in Fig.~\ref{fig:O_bow_scaling}(d). Further, we notice that if one performs the same scaling but for the first excited state with the same momentum quantum numbers as the ground state, a \textit{finite} (and with similar order of magnitude for the order parameter in the GS of the BOW phase shown in Fig.~\ref{fig:O_bow_scaling}) order parameter is obtained within this CDW phase in equilibrium.

Since we argue that the enhancement of the bond order by the pump is intrinsically related to the maximization of the overlap of the time-evolving wave-function with the first excited state, these results suggest that our pump protocol (which obeys translation-invariance) could possibly be associated with a transient enhancement of long-range BOW order. We further notice that other studies investigating the first excited state of variants of the extended Hubbard model [\href{https://doi.org/10.1143/JPSJ.78.044713}{55}], have argued that they manifest a formation of local bound singlet spin pairs, which one can face as the building blocks for the proper BOW order. What we advance with our analysis is that one instead observes the formation of full long range order in the problem and not the short range indication, as suggested by the \textit{local} bound singlet spin pairs.

Certainly, it will be of fundamental importance to confirm these results with techniques that go beyond the small system sizes attainable by exact diagonalization calculations; this will be left for future studies.

\end{document}